\documentstyle[preprint,revtex,eqsecnum]{aps}
\begin{document}
\draft
\baselineskip = 1.1\baselineskip
\begin{title}
{Critical exponents for $3D$ $O(n)$--symmetric \\
model with $n > 3$}
\end{title}
\author{S. A. Antonenko and A. I. Sokolov}
\begin{instit}
Department of Physical Electronics, Electrotechnical University, \\
Professor Popov Str. 5, St.Petersburg, 197376, Russia
\end{instit}
\begin{abstract}
Critical exponents for the $3D$ $O(n)$--symmetric model with $n > 3$
are estimated on the base of six--loop renormalisation--group (RG) 
expansions. Simple Pad$\acute {\rm e}$--Borel technique is used for 
resummation of RG series and Pad$\acute {\rm e}$ approximants 
$[ L/1 ]$ are shown to give rather good numerical results for all 
calculated quantities. For large $n$, the fixed point location $g_c$ 
and critical exponents are also determined directly from six--loop 
expansions, without addressing to resummation procedure. Analysis 
of numbers obtained shows that resummation becomes unnecessary when 
$n$ exceeds 28 provided an accuracy about 0.01 is adopted as 
satisfactory for $g_c$ and critical exponents. Further, 
results of the 
calculations performed are used to estimate numerical 
accuracy of the $1 \over n$--expansion. The same value, $n = 28$, is 
shown to play a role of lower boundary of the domain where this 
approximation provides high--precision estimates for critical 
exponents.
\end{abstract}
\narrowtext

\section{Introduction}
\label{sec:1}
Field--theoretical $3D$ $O(n)$--symmetric model with self--interaction
of $\lambda \varphi^4$ type is known to describe critical behaviour 
of many basic physical systems such as Ising ($n = 1$) and Heisenberg 
($n = 3$) ferromagnets, superfluid Bose--liquid ($n = 2$), polymers 
($n = 0$), etc. In 70--th Nickel, Meiron, and Baker, Jr. calculated 
all 2--point and 4--point Feynman graphs for this model up to 
six--loop order \cite{1} paving the way for obtaining perturbative 
expansions of unprecendented length for $\beta$--function and 
critical exponents. These expansions were then explicitly found and 
used, being resummed in various manners, to estimate the stable 
fixed point 
coordinate and numerical values of critical exponents \cite{2,3,4}. 
The values obtained are referred today as most accurate (canonical) 
numbers \cite{5}. 

Explicit expressions for RG functions and numerical estimates were 
presented in Refs. \cite{2,3,4} only for $n = 0, 1, 2, 3$. At the 
same time, it is desirable to have such results for $n > 3$. They are 
interesting from, at least, three points of view. First, there are 
numerous physical systems with many--component order parameters and 
these results may be relevant to their critical or effective critical 
behaviour (see, e.g. Refs. \cite{6,7,8}). Second, such calculations 
would enable one to clear up where resummation procedures applied to 
RG series become unnecessary, i.e. how large are the values of $n$ 
for which the theory may be thought as possessing a small 
parameter. And third, high--precision numerical estimates of critical 
exponents for $n \gg 1$ when compared with their counterparts given 
by $1 \over n$--expansion would provide an information about the 
numerical accuracy of this familiar approximation scheme. 

Below, six--loop perturbative expansions for $\beta$--function and 
critical exponents $\eta$ and $\gamma$ ($\gamma^{-1}$) are calculated 
for arbitrary $n$. The fixed point coordinate $g_c$ and critical 
exponents are estimated on the base of Pad$\acute {\rm e}$--Borel 
resummation procedure and a comparison of these numbers with those 
given by unresummed RG series and $1 \over n$--expansion is made. The 
outline of the paper is as follows. In Sec.~\ref{sec:2} the 
renormalisation scheme is formulated, RG expansions are written down, 
the resummation technique is described, and numerical results 
obtained are collected. In Sec.~\ref{sec:3} they are discussed along 
with their analogues resulting from unresummed six--loop series and 
$1 \over n$--expansion and corresponding inferences are presented. 
Section \ref{sec:4} contains conclusions. 

\section{RG series and numerical results}
\label{sec:2}
The Hamiltonian of the model to be studied reads: 
\begin{equation}
H = {1 \over 2} \int d^3 x \Bigl[ (\nabla \varphi_{\alpha})^2 + 
m_0^2 \varphi_{\alpha}^2 + {2 \over {4 !}} \lambda 
( \varphi_{\alpha}^2 )^2 \Bigr] \ \ , 
\label{eq:2.1}
\end{equation}
where $\varphi_{\alpha}$ is a vector order parameter field, 
$\alpha = 1, \ldots, n$, a bare mass squared $m_0^2$ being 
proportional to the deviation from the mean--field transition point. 

We calculate the $\beta$--function and critical exponents within a 
massive theory. The renormalized Green function $G_R (p, m, g)$ and 
four--point vertex function $\Gamma_R (p, m, g)$ are normalized at 
zero momenta in a conventional way: 
\begin{eqnarray}
G_R^{-1} (0, m, g) = m^2 \ \ , \qquad \ \nonumber \\
{{\partial G_R^{-1} (p, m, g)} \over {\partial p^2}}
\bigg\arrowvert_{p^2 = 0} = 1 \ \ , \label{eq:2.2} \\
\Gamma_R (0, m, g) = mg \ \ , \qquad \ \ \nonumber
\end{eqnarray}
with one extra condition being imposed on the $\varphi^2$ insertion: 
\begin{equation}
\Gamma_R^{1,2} (p, q, m, g)\Big\arrowvert_{p = q = 0} = 1 \ \ . 
\label{eq:2.3}
\end{equation}
Since combinatorial factors and momentum integrals for 2--point and 
4--point Feynman graphs are known \cite{1} the calculation of the 
$\beta$--function and critical exponents (anomalous dimensions) 
within six--loop approximation is straightforward (see, e.g. 
\cite{9}). The results are as follows: 
\FL
\begin{eqnarray}
\beta (g) = g - g^2 + {1 \over (n + 8)^2} \biggl( 6.07407408 n + 
28.14814815 \biggr) g^3 - {1 \over (n + 8)^3} \biggl( 1.34894276 n^2 
\qquad \ \ \ 
\nonumber \\
+ 54.94037698 n + 199.6404170 \biggr) g^4 + {1 \over (n + 8)^4} 
\biggl( - 0.15564589 n^3 + 35.82020378 n^2 \qquad 
\nonumber \\
+ 602.5212305 n + 1832.206732 \biggr) g^5 - {1 \over (n + 8)^5} 
\biggl( 0.05123618 n^4 + 3.23787620 n^3 \qquad \quad \ 
\nonumber \\
+ 668.5543368 n^2 + 7819.564764 n + 20770.17697 \biggr) g^6 + 
{1 \over (n + 8)^6} \biggl( - 0.02342417 n^5 \qquad 
\nonumber \\
- 1.07179839 n^4 + 265.8357032 n^3 + 12669.22119 n^2 + 
114181.4357 n \qquad \qquad \qquad \quad \ \ 
\nonumber \\
+ 271300.0372 \biggr) g^7 \ \ , \qquad \qquad \qquad \qquad \qquad 
\qquad \qquad \qquad \qquad \qquad \qquad \qquad \qquad \quad \ \ 
\label{eq:2.4}
\end{eqnarray}
\begin{eqnarray}
\eta (g) = {1 \over (n + 8)^2} \biggl( 0.2962962964 n + 0.5925925928 
\biggr) g^2 + {1 \over (n + 8)^3} \biggl( 0.0246840014 n^2 
\qquad \qquad \ \ 
\nonumber \\
+ 0.246840014 n 
+ 0.3949440224 \biggr) g^3 + {1 \over (n + 8)^4} \biggl( 
- 0.0042985626 n^3 \qquad \qquad \qquad \qquad 
\nonumber \\
+ 0.6679859202 n^2 + 4.609221057 n 
+ 6.512109933 \biggr) g^4 - {1 \over (n + 8)^5} \biggl( 
0.0065509222 n^4 \ \ \ \ 
\nonumber \\
- 0.1324510614 n^3 + 1.891139282 n^2 + 15.18809340 n 
+ 21.64720643 \biggr) g^5 \qquad \qquad \qquad 
\nonumber \\
+ {1 \over (n + 8)^6} \biggl( - 0.0055489202 n^5 - 0.0203994485 n^4 
+ 3.054030987 n^3 \qquad \qquad \qquad \qquad 
\nonumber \\
+ 64.07744656 n^2 + 300.7208933 n + 369.7130739 \biggr) g^6 \ \ , 
\qquad \qquad \qquad \qquad \qquad \qquad \ \ \ 
\label{eq:2.5}
\end{eqnarray}
\begin{eqnarray}
\gamma^{-1} (g) = 1 - {n + 2 \over 2 (n + 8)} g 
+ {n + 2 \over (n + 8)^2} g^2 
- {1 \over (n + 8)^3} \biggl( 0.8795588926 n^2 + 6.485476868 n 
\qquad \qquad 
\nonumber \\
+ 9.452718166 \biggr) g^3 + {1 \over (n + 8)^4} \biggl( 
- 0.1283321043 n^3 + 7.966740703 n^2 \qquad \qquad \qquad \quad \ 
\nonumber \\
+ 51.84421298 n + 70.79480631 \biggr) g^4 
- {1 \over (n + 8)^5} \biggl( 0.0490966058 n^4 \qquad \qquad 
\qquad \qquad \ 
\nonumber \\
+ 4.288152493 n^3 + 108.3618219 n^2 
+ 537.8136105 n + 675.6996077 \biggr) g^5 \qquad \qquad \quad \ \ 
\nonumber \\
+ {1 \over (n + 8)^6} \biggl( - 0.0259267945 n^5 - 1.618627843 n^4 
+ 85.54569746 n^3 \qquad \qquad \qquad \quad \ \ 
\nonumber \\
+ 1538.818235 n^2 + 6653.956526 n + 7862.074086 \biggr) g^6 \ \ . 
\qquad \qquad \qquad \qquad \qquad \qquad \ 
\label{eq:2.6}
\end{eqnarray}

These series are known to be divergent (asymptotic). To extract the 
physical information which they contain some resummation procedure 
should be employed. We use the Pad$\acute {\rm e}$--Borel method, 
i.e. construct Pad$\acute {\rm e}$ approximants $[ L/M ]$ for Borel 
transforms which are related to functions to be found (``sum of 
series'') by the formula 
\begin{equation}
f(x) = \sum_{k = 0}^{\infty} c_k x^k = \int\limits_0^{\infty} 
e^{-t} F(xt) dt \ \ , \label{eq:2.7}
\end{equation}
\begin{equation}
F(y) = \sum_{k = 0}^{\infty} {c_k \over {k!}} y^k \ \ , 
\label{eq:2.8}
\end{equation}
and then evaluate the integral (\ref{eq:2.7}) where series 
(\ref{eq:2.8}) possessing non--zero radii of convergence are 
replaced by corresponding Pad$\acute {\rm e}$ approximants. 

Starting from six--loop expansions available it is possible to 
construct different sets of Pad$\acute {\rm e}$ approximants: 
$[ L/1 ]$, $[ L - 1 /2 ]$, etc., where $L = 6$ for $\beta$--function 
and $L = 5$ for critical exponents. As we found, approximants 
\begin{equation}
[ L/1 ] = (1 + b_1 y)^{-1} \sum_{i = 0}^L a_i y^i \label{eq:2.8a}
\end{equation}
which generate following expressions for sums of the series 
\begin{eqnarray}
f(x) = z e^{-z} Ei(z) \sum_{i = 0}^L a_i (-b_1)^{-i} - 
\sum_{i = 1}^L 
a_i (-b_1)^{-i} \sum_{k = 0}^{i - 1} k! z^{-k} \ \ , \nonumber \\
z = - {1 \over {b_1 x}} \ \ , \qquad Ei(z) = \int\limits_{-\infty}^z 
e^t t^{-1} dt \qquad \qquad \qquad \qquad \ \ \ \ 
\label{eq:2.8b}
\end{eqnarray}
give the best results. They are presented in Table~\ref{tab:1}. 
The estimates for $\gamma$ and $\eta$ originate from series 
(\ref{eq:2.5}) and (\ref{eq:2.6}) while numerical values of critical 
exponents $\nu$, $\alpha$, and $\beta$ were determined by means of 
well--known scaling relations. The exponent $\gamma$ was calculated 
also via resummed RG expansion for the exponent $\eta_2 = (1 - 
\gamma) (2 - \eta) / \gamma$ and numbers were obtained which differ 
from those resulting from (\ref{eq:2.6}) by no more than 0.003; 
corresponding averages stand in Table~\ref{tab:1}. This table 
contains as well, for comparison, numerical results found earlier 
for $n = 0, 1, 2, 3$ on the base of higher--order RG expansions in 
3 and $4 - \epsilon$ dimensions using alternative resummation 
techniques \cite{2,3,10}. It is worthy to discuss these results 
along with ours in more detail. 

As we can see, there are small differences between our estimates 
and their counterparts obtained in Refs.~\cite{2,3} from $3D$ RG 
expansions of the same length. They are caused by use of different 
resummation procedures. Indeed, the authors of Refs.~\cite{2,3} 
employed the Borel--Leroy transformation 
\begin{equation}
f(x) = \int\limits_0^{\infty} t^B e^{-t} {\cal F}(xt) dt 
\label{eq:2.9}
\end{equation}
instead of Eq.~(\ref{eq:2.7}) in their calculations. The parameter 
$B$ was chosen to meet the known large--order behaviour of 
coefficients $c_k$ in perturbative expansions \cite{11,12}: 
\begin{equation}
c_k \sim  k! (- a)^k k^b \ \ , \qquad k \to \infty \ \ , 
\label{eq:2.10}
\end{equation}
where $a = 0.147774$ for the model (\ref{eq:2.1}) and $b$ is equal 
to $2 + {n \over 2}$, or $3 + {n \over 2}$, or $5 + {n \over 2}$ 
depending on the RG function expanded. We use much simpler method 
which ignores some part of information (\ref{eq:2.10}) but leads, 
nevertheless, to numerical results rather close to those given by 
more sophisticated techniques. It is not surprising since the main 
property of $c_k$~--~their factorial growth, is taken into account 
in our analysis while the rest of information about $c_k$ being 
incorporated enables one to reduce the apparent errors of estimation 
keeping the location of fixed point and critical exponents 
practically unchanged (see, e.g. Ref.~\cite{3} for detail). Dealing 
with simple Pad$\acute {\rm e}$ approximants $[ L/1 ]$, we avoid 
also, to a certain extent, the problem of poles. The point is 
that these approximants turn out to have no real and positive 
poles for $n < 38$ in the case of critical exponents and up to 
$n = 80$ for the $\beta$--function. That is why Table~\ref{tab:1} 
ends at $n = 32$. Since for $n = 0, 1, 2, 3$ our procedure gives 
critical exponents values which are almost identical to known 
high--precision estimates \cite{2,3,10}, we believe that the rest 
of the results listed in this table are also very close to exact 
numbers. 

\section{Large $n$ and $1 \over n$--expansion}
\label{sec:3}
How can we estimate $g_c$ and critical exponents for $n \agt 30$ ? 
It is well known (and clearly seen from 
Eqs.~(\ref{eq:2.4})--(\ref{eq:2.6})) that coefficients of RG 
expansions are decreasing when $n$ grows up. Hence, for large enough 
$n$ the theory should possess a true small parameter as, say, the 
quantum electrodynamics does. In such a case, all quantities of 
interest can be obtained directly from corresponding perturbative 
expansions, without addressing to resummation technique. To find the 
minimal value of $n$ which may be referred to as ``large enough'' we 
have calculated $g_c$ for $20 \le n \le 60$ using original and 
Pad$\acute {\rm e}$--Borel--resummed series (\ref{eq:2.4}). 
(It should be 
reminded that the approximant $[ 6/1 ]$ for the Borel transform of 
$\beta$--function has no dangerous poles within this segment.) 
The 
results are presented in Table~\ref{tab:2}. Values of $g_c$ given by 
these two approximations are seen to differ from each other by 
$0.9 \%$ for $n = 28$ and this difference diminishes rapidly with 
increasing $n$. So, if the accuracy of order of $1 \%$ for $g_c$ was 
adopted as satisfactory, the resummation of six--loop expansion for 
$\beta$--function becomes unnecessary when $n$ exceeds 28. The same 
turns out to be truth for the critical exponent $\gamma$ as is seen 
from Table~\ref{tab:3} (the first and the second lines).

For large $n$, another approximate method may be used to calculate 
critical exponents. We mean famous $1 \over n$--expansion. Within 
the second order in $1 \over n$ exponents $\gamma$ and $\eta$ are 
known to be \cite{13}:
\begin{equation}
\gamma = 2 - {24 \over \pi^2}{1 \over n} + {64 \over \pi^4} 
\Bigl( {44 \over 9} - \pi^2 \Bigr) {1 \over n^2} \ \ , 
\label{eq:3.1}
\end{equation}
\begin{equation}
\eta = {8 \over {3 \pi^2}}{1 \over n} - {512 \over {27 \pi^4}} 
{1 \over n^2} \ \ . \label{eq:3.2} 
\end{equation}
Series for other critical exponents are easily obtained via scaling 
relations.

It is interesting to evaluate the accuracy of numerical results 
given by $1 \over n$--expansion. We can get such an information 
comparing numbers
resulting from Eqs.~(\ref{eq:3.1}) and (\ref{eq:3.2}) 
for various $n$ with their counterparts obtained on the base of 
resummed ($n \le 32$) and unresummed ($n > 32$) six--loop RG series. 
On the other hand, this comparison would help us to determine the 
accuracy of the employed approximation itself in the limit 
$n \to \infty$ where $1 \over n$--expansion's results are exact. 

Corresponding estimates for exponent $\gamma$ are listed in 
Table~\ref{tab:3}. These numbers show that numerical accuracy of 
Eq.~(\ref{eq:3.1}) becomes better than $1 \%$ when $n$ exceeds 28. 
Values of $\eta$ given by six--loop RG series and Eq.~(\ref{eq:3.2}) 
are very small and not presented here. They differ from each other by 
approximately $10 \%$ for $n > 28$. Moreover, this discrepancy persists 
up to largest values of $n$ studied. It is not surprise. The point is 
that, for extremely large $n$, only leading terms in $n$ contribute 
to $\eta$ in each order in $g$. Since $g_c = 1 + O(n^{-1})$, $g_c$ 
should be put equal to unity within this limit. Hence, corresponding 
total contribution in the case of six--loop RG series may be found 
by summing of coefficients of all leading terms in Eq.~(\ref{eq:2.5}). 
Such a procedure gives $\eta = {0.30458 \over n}$, while the exact 
asymptotic expression resulting from Eq.~(\ref{eq:3.2}) is 
$\eta = {0.27019 \over n}$. So, the approximate asymptotic estimate 
for $\eta$ differs from the exact one by $13 \%$. This difference, 
however, practically doesn't influence upon numerical values of other 
critical exponents calculable by scaling relations since for $n > 28$ 
the exponent $\eta < 0.01$. 

We see that simple formulas (\ref{eq:3.1}) and (\ref{eq:3.2}) enable 
one to estimate all critical exponents for the model (\ref{eq:2.1}) 
with an accuracy of order of 0.01 provided $n \ge 28$. Moreover, for 
such $n$ second--order terms in these formulas may be, in fact, 
neglected since their contributions are very small.

\section{Conclusion}
\label{sec:4}
Critical exponents of the $3D$ $O(n)$--symmetric model have been 
estimated from six--loop RG series for $n > 3$. RG expansions have 
been resummed by means of simple Pad$\acute {\rm e}$--Borel technique 
and approximants $[ 6/1 ]$ ($\beta$--function) and $[ 5/1 ]$ 
(critical exponents) have been shown to provide rather good numerical 
results for all calculated quantities. It has been found that for 
$n \ge 28$ the theory may be thought as possessing a small parameter, 
i.e. the fixed point coordinate and critical exponents may be 
determined with errors about 0.01 or less directly from higher--order 
RG series, without use of resummation procedure. Numerical accuracy 
of the $1 \over n$--expansion has been also estimated. The same 
value, $n = 28$, has been shown to play a role of a lower boundary 
of the region where this approximation provides high--precision 
results for critical exponents. 

\acknowledgements
One of us (A.~I.~S.) cordially thanks B.~G.~Nickel for sending the 
Guelph report \cite{1} which was of key importance for completion 
of this work. We acknowledge also the support provided by the 
Russian Federation State Committee for Higher Education through 
Grant No.~94--7.17--351.

\widetext
\begin{table}
\caption{The stable fixed point location and critical exponents 
obtained within six--loop approximation using the
Pad$\acute {\rm e}$--Borel resummation technique.}
\begin{tabular}{ccccccc}
$n$ & $g_c$ & $\gamma$ & $\eta$ & $\nu$ & $\alpha$ & $\beta$ \\
\tableline
0 & 1.402 & 1.160 & 0.034 & 0.589 & 0.231 & 0.305 \\
\\
& 1.421$^{\dagger}$ & 1.161$^{\dagger}$ & 0.026$^{\dagger}$ & 
0.588$^{\dagger}$ & 0.236$^{\dagger}$ & 0.302$^{\dagger}$ \\
& 1.417$^{\ddagger}$ & 1.162$^{\ddagger}$ & 0.026$^{\ddagger}$ & 
0.588$^{\ddagger}$ & & 0.302$^{\ddagger}$ \\
\tableline
1 & 1.401 & 1.239 & 0.038 & 0.631 & 0.107 & 0.327 \\
\\
& 1.416$^{\dagger}$ & 1.241$^{\dagger}$ & 0.031$^{\dagger}$ & 
0.630$^{\dagger}$ & 0.110$^{\dagger}$ & 0.324$^{\dagger}$ \\
& 1.414$^{\ddagger}$ & 1.240$^{\ddagger}$ & 0.032$^{\ddagger}$ & 
0.630$^{\ddagger}$ & & 0.325$^{\ddagger}$ \\
& & & 0.035$^{\ast}$ & 0.628$^{\ast}$ & & \\
\tableline
2 & 1.394 & 1.315 & 0.039 & 0.670 & - 0.010 & 0.348 \\
\\
& 1.406$^{\dagger}$ & 1.316$^{\dagger}$ & 0.032$^{\dagger}$ & 
0.669$^{\dagger}$ & - 0.007$^{\dagger}$ & 0.346$^{\dagger}$ \\
& 1.405$^{\ddagger}$ & 1.316$^{\ddagger}$ & 0.034$^{\ddagger}$ & 
0.669$^{\ddagger}$ & & 0.346$^{\ddagger}$ \\
& & & 0.037$^{\ast}$ & 0.665$^{\ast}$ & & \\
\tableline
3 & 1.383 & 1.386 & 0.038 & 0.706 & - 0.117 & 0.366 \\
\\
& 1.392$^{\dagger}$ & 1.390$^{\dagger}$ & 0.031$^{\dagger}$ & 
0.705$^{\dagger}$ & - 0.115$^{\dagger}$ & 0.362$^{\dagger}$ \\
& 1.391$^{\ddagger}$ & 1.387$^{\ddagger}$ & 0.034$^{\ddagger}$ & 
0.705$^{\ddagger}$ & & 0.365$^{\ddagger}$ \\
& & & 0.037$^{\ast}$ & 0.698$^{\ast}$ & & \\
\tableline
4 & 1.369 & 1.449 & 0.036 & 0.738 & - 0.213 & 0.382 \\
\tableline
5 & 1.353 & 1.506 & 0.034 & 0.766 & - 0.297 & 0.396 \\
\tableline
6 & 1.336 & 1.556 & 0.031 & 0.790 & - 0.370 & 0.407 \\
\tableline
7 & 1.319 & 1.599 & 0.029 & 0.811 & - 0.434 & 0.417 \\
\tableline
8 & 1.303 & 1.637 & 0.027 & 0.830 & - 0.489 & 0.426 \\
\tableline
9 & 1.288 & 1.669 & 0.025 & 0.845 & - 0.536 & 0.433 \\
\tableline
10 & 1.274 & 1.697 & 0.024 & 0.859 & - 0.576 & 0.440 \\
\tableline
12 & 1.248 & 1.743 & 0.021 & 0.881 & - 0.643 & 0.450 \\
\tableline
14 & 1.226 & 1.779 & 0.019 & 0.898 & - 0.693 & 0.457 \\
\tableline
16 & 1.207 & 1.807 & 0.017 & 0.911 & - 0.732 & 0.463 \\
\tableline
18 & 1.191 & 1.829 & 0.015 & 0.921 & - 0.764 & 0.468 \\
\tableline
20 & 1.177 & 1.847 & 0.014 & 0.930 & - 0.789 & 0.471 \\
\tableline
24 & 1.154 & 1.874 & 0.012 & 0.942 & - 0.827 & 0.477 \\
\tableline
28 & 1.136 & 1.893 & 0.010 & 0.951 & - 0.854 & 0.481 \\
\tableline
32 & 1.122 & 1.908 & 0.009 & 0.958 & - 0.875 & 0.483 \\
\end{tabular}
\label{tab:1}
\tablenotes{$^{\dagger}$~Quoted from Ref.~\cite{3}. \\
$^{\ddagger}$~Quoted from Ref.~\cite{2}. \\
$^{\ast}$~Quoted from Ref.~\cite{10}.}
\end{table}
\begin{table}
\caption{Coordinates of the fixed point obtained from 
Eq.~(\ref{eq:2.4})with use of 
Pad$\acute {\rm e}$--Borel resummation procedure (PB) and by 
direct summation (DS).} 
\begin{tabular}{cccccccccc}
$n$ & & 20 & 24 & 28 & 32 & 36 & 40 & 50 & 60 \\
\tableline
$g_c$ & (DS) & 1.2184 & 1.1725 & 1.1458 & 1.1273 & 1.1134 & 1.1025 & 
1.0830 & 1.0699 \\
& (PB) & 1.1768 & 1.1538 & 1.1359 & 1.1216 & 1.1099 & 1.1003 & 
1.0822 & 1.0696 \\
\end{tabular}
\label{tab:2}
\end{table}
\begin{table}
\caption{Values of the critical exponent $\gamma$ obtained by 
direct summation of the RG expansion (DS), 
by means of Pad$\acute {\rm e}$--Borel resummation 
technique (PB) and from $1 \over n$--expansion  
(Eq.~(\ref{eq:3.1})).}
\begin{tabular}{ccccccccccc}
$n$ & 20 & 24 & 28 & 32 & 36 & 40 & 50 & 70 & 100 & 500 \\
\tableline
(DS) & 1.8990 & 1.8991 & 1.9075 & 1.9165 & 1.9245 & 1.9314 & 
1.9447 & 1.9606 & 1.9725 & 1.9946 \\
(PB) & 1.8466 & 1.8737 & 1.8932 & 1.9078 & 1.9222 & & & & & \\
$1 \over n$ & 1.8702 & 1.8930 & 1.9090 & 1.9208 & 1.9299 & 1.9372 
& 1.9501 & 1.9646 & 1.9754 & 1.9951 \\
\end{tabular}
\label{tab:3}
\end{table}

\end{document}